\documentclass[aps,prd,twocolumn,floats,showpacs]{revtex4}
\usepackage{graphics,epsfig,color} 
\usepackage{amsmath,amssymb}

\def\be{\begin{equation}}
\def\ee{\end{equation}}
\def\bea{\begin{eqnarray}}
\def\eea{\end{eqnarray}}

\begin{document}


\title{Extended $f\left(R,L_m \right)$ gravity with generalized scalar field and kinetic term dependences}

\author{Tiberiu Harko$^1$}\email{t.harko@ucl.ac.uk}
\author{Francisco S.N.~Lobo$^{2}$}\email{flobo@cii.fc.ul.pt}
\author{Olivier Minazzoli$^{3}$}\email{ominazzo@caltech.edu}

\affiliation{$^1$Department of Mathematics, University College London, Gower Street, London WC1E 6BT, United Kingdom}
\affiliation{$^2$Centro de Astronomia e Astrof\'{\i}sica da Universidade de Lisboa, Campo Grande, Ed. C8 1749-016 Lisboa, Portugal}

\affiliation{$^3$Jet Propulsion Laboratory, California Institute of Technology,\\
4800 Oak Grove Drive, Pasadena, CA 91109-0899, USA}

\date{\today}

\begin{abstract}

We generalize previous work by considering a novel gravitational model with an action given by an arbitrary function of the Ricci scalar, the matter Lagrangian density, a scalar field and a kinetic term constructed from the gradients of the scalar field, respectively.  The gravitational field equations in the metric formalism are obtained, as well as the equations of motion for test particles, which follow from the covariant divergence of the
stress-energy tensor. Specific models with a nonminimal coupling between the scalar field and the matter Lagrangian are further explored. We emphasize that these models are extremely useful for describing an interaction between dark energy and dark matter, and for explaining the late-time cosmic acceleration.
\end{abstract}


\maketitle

\section{Introduction}

Scalar fields play a fundamental role in the cosmological description of our Universe \cite{Fa04}. One of the first major extensions of general relativity, proposed by Dicke and Brans \cite{Di61, Br1, Br2}, conjectured that  ``Mach's principle'' might lead to a dependence of the local Newtonian gravitational constant, $G$, since in most cosmological models the total mass $M$ and radius of curvature of the Universe $a(t)$ are related by an equation of the form $G^{-1}\sim M/a(t)c^2$. Consequently, in the variational principle of general relativity \cite{Di61, Br1, Br2}, it was proposed to substitute $G^{-1}$ with a scalar field $\phi$, and to also add to the action the kinetic energy corresponding to $\phi $. Therefore, the variational principle of the Brans-Dicke theory can be formulated as $\delta \int{\left(\phi R +L_m-\omega \nabla _{\lambda }\phi \nabla ^{\lambda }\phi /\phi \right)\sqrt{-g}d^4x}=0$, where $R$ is the scalar curvature, $L_m$ is the matter Lagrangian, and $\omega $ is a coupling parameter.
The scalar-tensor gravitational models have been intensively investigated, and can be considered as a valid approach in explaining the recent accelerated expansion of the Universe, inferred from the Type Ia supernova observations \cite{Ri98}. According to standard general relativity, the observed late-time cosmic acceleration can be successfully explained by introducing either a fundamental cosmological constant \cite{bianchiNATURE10} -- which would represent an intrinsic curvature of space-time -- or a dark energy -- which would \textit{mimic} a cosmological constant (at least during the late stage of the cosmological evolution), as the concordance of observations are still in favor of the $\lambda$-CDM standard model \cite{PeRa03}, where $\lambda$ is a constant. One of the currently main dark energy scenarios is based on the so-called quintessence, introduced in \cite{quint0} and \cite{quint}, where dark energy corresponds to a new scalar particle $\Phi$.

After the Brans-Dicke proposal, other forms of scalar-tensor theoretical models were investigated. Cosmologies where the Planck length is not a fundamental constant but rather evolves with time were considered in \cite{Wet2}, and it was shown that the dynamics which should be responsible for today's tiny value of this length scale are governed by the effective potential of a Brans-Dicke type theory. One can also extend the coupling of the scalar field and curvature to a nonminimal coupling, provided by $F(\phi )R$, where, for example, $F(\phi ) = 1-\xi \phi ^2$, and $\xi $ is a coupling constant \cite{Be}. In these types of models non-minimally coupled Higgs field with a large coupling $\xi $ might give rise to a successful inflation \cite{Be}, which is otherwise very difficult to be achieved \cite{Ket}. One of the positive features of Brans-Dicke-like theories is that they seem to be generically driven toward general relativity during the cosmological evolution of the Universe \cite{STT2GR}.

Couplings between the scalar and the matter fields have been investigated as well \cite{gasperiniPRD02,damourPRD96,dilaton2EP,armendarizPRD02,EP,int,NMC,min, Cota}. Indeed, such couplings generically appear in Kaluza-Klein theories with compactified dimensions \cite{KK} or in the low energy effective limit of string theories \cite{gasperiniPRD02,damourPRD96,dilaton2EP}. In the latter context, some authors proposed that the dilaton could be a good candidate for the quintessence \cite{gasperiniPRD02} or the inflaton \cite{damourPRD96}. But even from a phenomelogical point of view, it has been argued that specific restrictions such as gauge and diffeomorphism invariances essentially single out a particular set of effective theories which turns out to be Brans-Dicke-like theories with scalar/matter coupling \cite{armendarizPRD02}. A good feature of such theories is that under various assumptions -- and similarly to Brans-Dicke-like theories without scalar/matter coupling -- they seem to be driven toward a weak coupling through cosmological evolution \cite{dilaton2EP}. Therefore, they seem to be able to explain the current tight constraints on the equivalence principle without fine-tuning parameters \cite{EP}.

More recently, it has been argued that a scalar/matter coupling could be responsible for a dependency of the effective mass of the scalar-field upon the local matter density \cite{motaPLB04}. Such a scalar-field has been dubbed \textit{chameleon} as ``in regions of high density, the [scalar-field] chameleon blends with its environment and becomes essentially invisible to searches for EP violation and fifth force'' \cite{khouryPRD04}.

The possibility of a nonminimal coupling between matter and geometry, somehow similar to the coupling between the scalar field and curvature in the Brans-Dicke theory, with the matter Lagrangian playing the role of $\phi $,  was considered in \cite{Ko06,Bertolami:2007gv}, in the framework of the so-called $f(R)$ models of gravity \cite{Bu70, Carroll:2003wy, SoFa08}, and further extended in \cite{Bertolami:2007vu}.
An extension of the $f(R)$ gravity model, called $f(R,\phi )$ gravity was also proposed in \cite{Hwang}, and further discussed and generalized in \cite{Mat}. The action considered for the $f(R,\phi )$ gravity model is given by $S=\int{\left[f(R,\phi )/2-\omega (\phi )\nabla _{\mu }\phi \nabla ^{\mu }\phi -V(\phi )+\beta L_m\right]\sqrt{-g}d^4x/\beta }$, where $\beta $ is a constant.
In this context, the maximal extension of the standard Hilbert-Einstein action of general relativity, $S=\int{\left(R/16\pi G+L_m\right)\sqrt{-g}d^4x}$, was considered in \cite{HaLo08}, where the $f(R)$ type gravity models were maximally generalized by assuming that the gravitational Lagrangian is given by an arbitrary function of the Ricci scalar $R$ and of the matter Lagrangian $L_m$, so that $S=\int{f\left(R,L_m\right)\sqrt{-g}d^4x}$.

It is the purpose of the present paper to consider a generalized $f\left(R,L_m\right)$ type gravity model, in which, beyond the ordinary matter, described by its thermodynamic energy density $\rho $ and pressure $p$, the Universe is filled with a scalar field $\phi $. Consequently, in the gravitational action the scalar field, as well as its kinetic energy, $\nabla _{\mu }\phi \nabla ^{\mu }\phi $ are also present, and the Lagrangian takes the form $L_{grav}=f\left(R,L_m, \phi, \nabla _{\mu }\phi \nabla ^{\mu }\phi \right)$, where $f$ is an arbitrary function.  Therefore, in these models an explicit nonminimal coupling between matter and the scalar field is also allowed. As a first step in our study we obtain the gravitational field equations in the metric formalism, and the equations of motion for test particles, which follow from the covariant divergence of the stress-energy tensor.
Some specific models with nonminimal scalar field-matter coupling are further explored.

\section{Gravitational field equations}\label{sect2}

We assume that the action for the modified theories of gravity, considered in this work, is given by
\begin{equation}\label{1}
S=\int f\left(R,L_m, \phi ,g^{\mu \nu}\nabla _{\mu }\phi \nabla _{\nu }\phi \right) \sqrt{-g}\;d^{4}x,
\end{equation}
where $\sqrt{-g}$ is the determinant of the metric tensor $g_{\mu \nu}$, and $f\left(R,L_m, \phi, g^{\mu \nu}\nabla _{\mu }\phi \nabla _{\nu }\phi \right)$ is an arbitrary function of the Ricci scalar $R$, the matter Lagrangian density, $L_{m}$, a scalar field $\phi $, and the gradients constructed from the scalar field, respectively. The only restriction on the function $f$ is  to be an analytical function of $R$, $L_m$, $\phi $, and of the scalar field kinetic energy, respectively, that is, $f$ must possess a Taylor series expansion about any point.

We define the ``reduced'' stress-energy tensor as \cite{LaLi}
\begin{equation}
\tau _{\mu \nu }=-\frac{2}{\sqrt{-g}}\frac{\delta \left( \sqrt{-g}L_{m}\right)}{
\delta g^{\mu \nu }}\,.
\end{equation}
By assuming that the matter Lagrangian density $L_{m}$ depends only
on the metric tensor components $g_{\mu \nu }$, and not on its derivatives,
we obtain the stress-energy tensor as
\begin{equation}\label{en1}
\tau _{\mu \nu }=g_{\mu \nu }L_{m}-2\frac{\delta L_{m}}{\delta g^{\mu
\nu }}.
\end{equation}
Furthermore, we assume that the scalar field $\phi $ is independent of the metric, i.e., $\delta \phi /\delta g^{\mu \nu }\equiv 0$. In the following we will denote, for simplicity,
$\left(\nabla \phi\right)^2=g^{\mu \nu}\nabla _{\mu }\phi \nabla _{\nu }\phi $.

By varying the action $S$ of the gravitational field with respect to the
metric tensor components $g^{\mu \nu }$, we obtain
the field equations of the $f\left[ R,L_{m}, \phi , \left(\nabla \phi \right)^2\right] $ gravitational model as
\begin{eqnarray}\label{field}
f_{R} R_{\mu \nu }+\left( g_{\mu \nu }\nabla _{\lambda }\nabla^{\lambda }
-\nabla
_{\mu }\nabla _{\nu }\right)f_R
- \frac{1}{2}\left(
f -f_{L_{m}}L_m \right)g_{\mu \nu }
    \nonumber  \\
=\frac{1}{2}%
f_{L_{m}} \tau _{\mu \nu }-f_{\left(\nabla \phi \right)^2 }\nabla _{\mu }\phi \nabla _{\nu }\phi ,\nonumber\\
\end{eqnarray}
where the subscript of $f$ denotes a partial derivative with respect to the arguments, i.e., $f_R=\partial f/\partial R$, $f_{L_m}=\partial f/\partial L_m$, $f_{\left(\nabla \phi \right)^2 }=\partial f/\partial \left(\nabla \phi \right)^2$.

Now varying the action with respect to $\phi $, provides the following evolution equation for the scalar field
\begin{equation}
\square _{\left( \nabla \phi \right) ^{2}}\phi =\frac{1}{2}f_{\phi },
\end{equation}
where $f_\phi=\partial f/\partial \phi$ and
\begin{equation}\label{2}
\square _{\left( \nabla \phi \right) ^{2}}=\frac{1}{\sqrt{-g}}\frac{\partial
}{\partial x^{\mu }}\left[ f_{\left( \nabla \phi \right) ^{2}}\sqrt{-g}%
g^{\mu \nu }\frac{\partial }{\partial x^{\nu }}\right] ,
\end{equation}
is the generalized D'Alembert  operator of $f\left(R,L_m, \phi, \nabla _{\mu }\phi \nabla ^{\mu }\phi \right)$ gravity.

The contraction of Eq.~(\ref{field}) provides the following relation between the Ricci scalar $R$, the matter Lagrangian density $L_{m}$, the derivatives of the scalar field, and the trace $\tau =\tau _{\mu }^{\mu }$ of the ``reduced'' stress-energy tensor,
\begin{eqnarray}
f_{R} R+3\nabla _{\mu }\nabla ^{\mu } f_{R} -2\left(
f-f_{L_{m}}L_{m}\right)=
   \nonumber\\
\frac{1}{2}f_{L_{m}} \tau -f_{\left( \nabla \phi \right) ^{2}}\nabla _{\mu }\phi \nabla ^{\mu }\phi .  \label{contr}
\end{eqnarray}

By taking the covariant divergence of Eq.~(\ref{field}), we obtain for the covariant divergence of the ``reduced''  stress-energy tensor the following expression
\begin{eqnarray}
&&\frac{1}{2} \nabla^\sigma \left(f_{L_m} \tau_{\mu \sigma}\right) = \frac{1}{2} \left( L_m \nabla_\mu f_{L_m}-f_\phi \nabla_\mu \phi \right)\nonumber\\
&&+f_{\left(\nabla \phi \right)^2}\nabla_\mu \phi \nabla_\sigma \nabla^\sigma \phi+
\nabla_\mu \phi \nabla^\sigma \phi \nabla_\sigma f_{\left(\nabla \phi \right)^2}.
\end{eqnarray}

This relationship was deduced by taking into account the following mathematical identities
\bea
\nabla ^{\mu }R_{\mu \nu }=\frac{1}{2} \nabla _{\nu }R \,,
\left( \nabla _{\nu }\square
-\square \nabla _{\nu }\right) f_{R}=-\left( \nabla ^{\mu }f_{R}\right)R_{\mu \nu } ,
\eea
and we have used the fact that we consider torsion-free space-times such that
$\left[ \nabla_\sigma \nabla_\epsilon - \nabla_\epsilon \nabla_\sigma \right] \psi =0$,
where $\psi$ is any scalar-field. Now, using Eq.~(\ref{2}) we get
\begin{eqnarray}
\nabla^\sigma \left(f_{L_m} \tau_{\mu \sigma} \right) = L_m \nabla_\mu f_{L_m}.
\end{eqnarray}

For $\phi \equiv 0$, Eqs. (\ref{field}) reduce to the field equations of the $f\left(R,L_m\right)$ model considered in \cite{HaLo08}. For $\phi \neq 0$, one recovers the good conservation equations for either general relativity and Brans-Dicke-like scalar-tensor theories (with and without scalar/matter coupling \cite{min}). For instance, the total Lagrangian of the simplest matter-scalar field-gravitational field theory, with scalar field kinetic term and a self-interacting potential $V(\phi)$ corresponds to the choice
\begin{equation}
f=\frac{R}{2}+L_{m}+\frac{\lambda }{2} g^{\mu \nu }\nabla _{\mu }\phi \nabla _{\nu }\phi +V(\phi ),
\end{equation}
 where $\lambda $ is a constant. The corresponding field equations can be immediately obtained from Eqs.~(\ref{field}) as
 \bea
&& R_{\mu \nu }-\frac{1}{2}Rg_{\mu \nu }=\tau _{\mu \nu }-\lambda \nabla _{\mu }\phi \nabla _{\nu }\phi
 + \nonumber\\
 &&\left[\frac{\lambda}{2} g^{\alpha \beta }\nabla _{\alpha }\phi \nabla _{\beta }\phi +V(\phi)\right]g_{\mu \nu }.
 \eea
The scalar field satisfies the evolution equation
\begin{equation}
\frac{1}{\sqrt{-g}}\frac{\partial }{\partial x^{\mu}}\left[\sqrt{-g}g^{\mu \nu }\frac{\partial \phi }{\partial x^{\nu }}\right]=\frac{1}{\lambda }\frac{dV(\phi )}{d\phi },
\end{equation}
while the stress-energy tensor obeys the conservation equation $\nabla ^{\sigma }\tau _{\mu \sigma }=0$.


\section{Models with nonminimal matter-scalar field coupling}\label{sect3_1}

As an example of the application of the formalism developed in the previous section, we consider a simple phenomenological model, in which a scalar field is non-minimally coupled to pressure-less matter with {\it rest mass density } $\rho $. For the action of the system we consider
\be
S=\int{\left[\frac{R}{2}-F(\phi )\rho +\lambda g^{\mu \nu }\nabla _{\mu }\phi \nabla _{\nu }\phi \right]\sqrt{-g}d^4x},
\ee
where $F(\phi )$ is an arbitrary function of the scalar field that couples non-minimally to ordinary matter. The field equations for this model are given by
\bea
R_{\mu \nu }-\frac{1}{2}Rg_{\mu \nu }&=&F(\phi )\rho u_{\mu }u_{\nu }
   \nonumber\\
&&\hspace{-1cm}+2\lambda \left[\nabla _{\mu }\phi \nabla _{\nu }\phi -\frac{1}{2}g_{\mu \nu }\nabla _{\alpha }\phi \nabla ^{\alpha }\phi \right],
\eea
where $u^{\mu }$ is the four-velocity of the matter fluid. The scalar field satisfies the evolution equation
\be \label{square1}
\square \phi =-\frac{1}{2\lambda }\frac{dF(\phi)}{d\phi }\rho,
\ee
where $\square $ is the usual d'Alembert operator defined in a curved space. The total stress-energy tensor of the scalar field-matter system is given by
\be
T_{\mu \nu }=F(\phi )\rho u_{\mu }u_{\nu }+ 2\lambda \left[\nabla _{\mu }\phi \nabla _{\nu }\phi -\frac{1}{2}g_{\mu \nu }\nabla _{\alpha }\phi \nabla ^{\alpha }\phi \right].
\ee

Through the Bianchi identities, the covariant divergence of $T^{\mu \nu }$ must be zero, that is, $\nabla _{\nu }T^{\mu \nu }=0$. In the following, we assume that the mass density current is conserved, i.e.,
$\nabla _{\nu }\left(\rho u^{\nu }\right)=0$. Using the latter condition, and the mathematical identity given by $\left[ \nabla_\sigma \nabla_\epsilon - \nabla_\epsilon \nabla_\sigma \right] \psi =0$, we obtain first
\be
F(\phi )\rho \left( u^{\nu }\nabla _{\nu }u^{\mu}\right)+\rho u^{\mu }u^{\nu }\frac{dF(\phi )}{d\phi }\nabla _{\nu }\phi +2\lambda \left(\nabla ^{\mu }\phi \right)\square \phi =0.
\ee
By eliminating the term $\square \phi $ with the help of Eq.~(\ref{square1}), we obtain
\be
u^{\nu }\nabla _{\nu }u^{\mu }+\left[\frac{d}{d\phi }\ln F(\phi )\right]\left(u^{\mu }u^{\nu }\nabla _{\nu }\phi -\nabla ^{\mu }\phi \right)=0.
\ee

Using the identity $u^{\nu }\nabla _{\nu }u^{\mu }\equiv \frac{d^2x^{\mu }}{ds^2}+\Gamma _{\alpha \beta }^{\mu }u^{\alpha }u^{\beta },$
where $\Gamma _{\alpha \beta }^{\mu }$ are the Christoffel symbols corresponding to the metric, the equation of motion of the test particles
non-minimally coupled to an arbitrary scalar field takes the form
\be\label{20}
\frac{d^2x^{\mu }}{ds^2}+\Gamma _{\alpha \beta }^{\mu }u^{\alpha }u^{\beta }+ \left[\frac{d}{d\phi }\ln F(\phi )\right]\left(u^{\mu }u^{\nu }\nabla _{\nu }\phi -\nabla ^{\mu }\phi \right)=0.
\ee

A particular model can be obtained by assuming that $F(\phi )$ is given by a  linear function,
\be
F(\phi )=\frac{\Lambda +1}{2}\left[1+\frac{1}{2}\left(\Lambda -1\right)\phi \right],
\ee
where $\Lambda $ is a constant. Then the equation of motion becomes
\be
\frac{d^2x^{\mu }}{ds^2}+\Gamma _{\alpha \beta }^{\mu }u^{\alpha }u^{\beta }+\left(u^{\mu }u^{\nu }-g^{\mu \nu }\right)\nabla _{\nu }\ln\left[1+\frac{\Lambda -1}{2}\phi \right]=0.
\ee

In order to simplify the field equations we adopt for $\lambda $ the value $\lambda =-\left(\Lambda ^2-1\right)/8$. Then Eq.~(\ref{square1}), determining the scalar field, takes the simple form
$\square \phi =\rho$.

The gravitational field equations take the form
\be
R_{\mu \nu}-\frac{1}{2}g_{\mu \nu }R=\frac{\Lambda +1}{2}T_{\mu \nu },
\ee
with the total stress-energy tensor given by
\bea
T_{\mu \nu }&=&\left[1+\frac{\Lambda -1}{2}\phi \right]\rho u_{\mu }u_{\nu }
 \nonumber\\
&&-\frac{\Lambda -1}{2}\left[\nabla _{\mu }\phi \nabla _{\nu }\phi -\frac{1}{2}g_{\mu \nu }\nabla _{\alpha }\phi \nabla ^{\alpha }\phi \right].
\eea
For $\Lambda =1$ we reobtain the general relativistic model for dust. Other possible choices of the function $F(\phi)$, such as $F(\phi )=\exp (\phi )$, can be discussed in a similar way.

A more general model can be obtained by adopting for the matter Lagrangian the general expression \cite{Fock, Ha10,Min}
\be\label{lagr}
L_m=-\left[\rho +\rho \int{\frac{dp(\rho )}{\rho }}-p(\rho )\right],
\ee
where $\rho $ is {\it the rest-mass energy density}, $p$ is the thermodynamic pressure, which, by assumption, satisfies a barotropic equation of state, $p=p\left(\rho \right)$.  By assuming that the matter Lagrangian does not depend on the derivatives of the metric, and that the particle matter fluid current is conserved [$\nabla _{\nu }\left(\rho u^{\nu }\right)=0$],  the Lagrangian given by Eq.~(\ref{lagr}) is the unique matter Lagrangian that can be constructed from the thermodynamic parameters of the fluid, and it is valid for all gravitational theories satisfying the two previously mentioned conditions  \cite{Min}.

The gravitational field equations and the equation describing the
matter-scalar field coupling are given by
\be
R_{\mu \nu }-\frac{1}{2}g_{\mu \nu }R=F(\phi ) ~\epsilon ~u_{\mu }u_{\nu }-p g_{\mu \nu }+\lambda Q_{\mu \nu},
\ee
   \be
\square \phi =\frac{1}{2\lambda }\frac{dF(\phi )}{d\phi }~\epsilon,
 \ee
where
$Q_{\mu \nu}=\nabla _{\mu }\phi \nabla _{\nu }\phi -\frac{1}{2}\nabla _{\lambda }\phi \nabla ^{\lambda }\phi g_{\mu \nu }$,
and where the total energy density is
$\epsilon =\rho+\rho \int{dp/\rho }-p$ \cite{Fock,Min}.
With the use of the conservation equation
$\nabla _{\nu }\left(\rho u^{\nu }\right)=0$,
one obtains the equation of motion of massive test particles as
\be\label{eqmot}
\frac{d^2x^{\mu }}{ds^2}+\Gamma _{\alpha \beta }^{\mu }u^{\alpha }u^{\beta }+\left(u^{\mu }u^{\nu }-g^{\mu \nu }\right)\nabla _{\nu }\ln\left[1+\int{\frac{dp}{\rho }} \right]=0,
\ee
The equation of motion (\ref{eqmot}) can also be derived from the variational principle $\delta \int{\sqrt{1+\int{dp/\rho}}\sqrt{g_{\mu \nu }u^{\mu }u^{\nu}}ds}=0$. Models with scalar field-matter coupling were considered in the framework of the Brans-Dicke theory \cite{NMC}, with the action of the model given by $S=\int{\left[\phi R/2+\left(\omega /\phi \right)\nabla _{\mu }\phi \nabla ^{\mu }\phi +F(\phi )L_m\right]\sqrt{-g}d^4x}$. Such models can give rise to a late time accelerated expansion of the Universe for very high values of the Brans-Dicke parameter $\omega $. Other models with interacting scalar field and  matter have been considered in \cite{int}. We emphasize that the gravitational theory considered in this work generalizes all of the above models.

\section{Discussion and conclusion}
In the present paper we have presented a novel gravitational theory where the Lagrangian is given by an arbitrary function of the Ricci scalar, matter Lagrangian, a scalar field and its kinetic term, respectively. The field equations for this model were obtained for the general case, and the divergence of the stress-energy tensor has been computed. Our model unifies into a single  mathematical formalism  all the known modifications of standard general relativity (Brans-Dicke model, standard scalar field and quintessence models, $f(R,L_m)$ gravity etc.). Moreover, this approach opens the possibility of a systematic study of the maximal extensions of the classical Hilbert-Einstein Lagrangian in the presence of scalar fields, thus providing a theoretical tool for exploring the limits of the presently known geometrical gravitational theories. Specific models with a nonminimal interaction between the scalar field and ordinary matter were also explored.

In this context, an interesting application of the formalism, outlined in this work, it to the recently proposed cosmological model, denoted as {\it growing neutrino quintessence model} \cite{gq} (and further developed in \cite{Ai}). In this model, dark energy, described by a scalar field (cosmon), evolves dynamically and interacts with the cosmological neutrino backgrounds. The coupling between dark energy and neutrinos (described by the Dirac equation with scalar field dependent neutrino masses) provides a solution to the coincidence problem, and explains the small value of the cosmological constant. A coupling between the cosmon and the neutrinos leads to an exchange of energy and momentum, and to the non-conservation of the individual stress-energy tensors. Note that the geodesic equations of motion are also modified due to the presence of a cosmon mediated fifth force and of a velocity dependent force. In addition to this, the formation of large-scale neutrino structures, the local variations in the dark energy and the backreaction on the background evolution were further analysed by using a relativistic N-body treatment of the neutrinos, combined with the computation of the local quintessence field.
We emphasize that the scalar field and the matter-field coupling sectors of the growing quintessence model are also particular cases of the general framework introduced in the  present paper. In particular, according to the growing neutrino quintessence model, the equation of motion of the neutrino is given by $du^{\alpha }/ds+\Gamma _{\mu \nu }^{\alpha }u^{\mu }u^{\nu }+\beta u^{\lambda }u^{\alpha }\nabla _{\lambda }\varphi-\beta \nabla ^{\alpha }\varphi =0$, where $\beta =-d\ln m_{\nu}/d\varphi$, and with $m_{\nu}$ the (scalar field dependent) neutrino mass \cite{Ai}. Due to the cosmon-neutrino coupling there is a deviation from the geodesic motion. Note that the equation of motion of the neutrino in this model is identical with the general Eq.~(\ref{20}), if we interpret the function $F\left(\phi \right)$ in terms of the neutrino mass.

In conclusion, the general formalism outlined in this work can be extremely useful in a variety of scenarios, such as, in describing the interaction between dark energy, modeled as a scalar field, and dark matter, or ordinary matter (neutrinos), with or without pressure, matter-scalar field interactions in inflation, as well as in the study of the interactions of the scalar field (representing dark matter and/or dark energy) and the electromagnetic component in the very early Universe.  Moreover, they can provide a realistic description of the late expansion of the Universe, where a possible interaction between ordinary matter and dark energy cannot be excluded {\it a priori}.
Work along these lines is presently underway and the cosmological consequences of the present theory will also be investigated in detail in a future publication.


\section*{Acknowledgments}
 FSNL acknowledges financial support of the Funda\c{c}\~{a}o para a Ci\^{e}ncia e Tecnologia through the grants CERN/FP/123615/2011 and CERN/FP/123618/2011.
OM was supported by an appointment to the NASA Postdoctoral Program at the Jet Propulsion Laboratory, California Institute of Technology, administered by Oak Ridge Associated Universities through a contract with NASA.

\end{document}